# Wannier Center Analysis on Possible Three-Dimensional Topological Phases in $\alpha$-Type Layered Organic Conductors

Toshihito Osada

*Department of Applied Physics, The University of Tokyo,*

*7-3-1 Hongo, Tokyo 113-8656, Japan.*

Topological features of possible three-dimensional (3D) states in $\alpha$-type layered organic conductors are investigated within a unified framework based on Wannier charge centers (WCCs), aiming to identify their actual topological states. Among the 3D Dirac/Weyl semimetal states of multilayered $\alpha$-(ET)$_2$I$_3$, the type-I Dirac semimetal state, induced by interlayer spin-orbit coupling (SOC), most effectively explains the observed chiral transport phenomena attributed to the chiral magnetic effect, which originates from the spiral structures of the WCC sheets. In multilayered $\alpha$-(BETS)$_2$I$_3$, a 3D weak topological insulator (TI) state consistently emerges, irrespective of the presence of interlayer SOC and/or inversion symmetry breaking. The strong TI state suggested by experimental observations appears unlikely to be realized.

The electronic states of $\alpha$-type layered organic conductors, $\alpha$-$(ET)_2I_3$ (short for $\alpha$-$(BEDT\text{-}TTF)_2I_3$) and $\alpha$-$(BETS)_2I_3$, have traditionally been treated as two-dimensional (2D) massless and massive Dirac fermion (DF) systems, with the small interlayer coupling neglected [1-7]. However, three-dimensional (3D) topological properties have recently observed at sufficiently low temperatures where $k_B T$ becomes smaller than the interlayer transfer energy. This indicates a dimensional crossover to 3D topological states in low-temperature regime. Although a simple stacking of 2D massless DF systems is expected to result in a 3D nodal-line semimetal, a 3D Dirac/Weyl semimetal state has been suggested in $\alpha$-$(ET)_2I_3$ under pressure, based on experimental observations of negative transverse magnetoresistance and planar Hall effect [8, 9], both attributed to the chiral magnetic effect (CME) [10]. In contrast, in the $\alpha$-$(BETS)_2I_3$, surface transport across the entire crystal surface, indicative of a 3D strong topological insulator (TI), has been reported [11], although a simple stacking of 2D TIs would typically yield a 3D weak TI. Thus, $\alpha$-type organic conductors provide a new platform for exploring topological physics.

In our previous work [12], we investigated possible topological states in multilayered $\alpha$-$(ET)_2I_3$ and $\alpha$-$(BETS)_2I_3$ under time-reversal symmetry (TRS), taking into account interlayer transfers with spin-orbit coupling (SOC) and the breaking of space-inversion symmetry (SIS), in order to explain the observed dimensional crossover. We pointed out that interlayer SOC, originating from the potential of the $I_3^-$ anion, may not be negligible in $\alpha$-type conductors, since electrons must traverse the $I_3^-$ anion layer during interlayer hopping. In $\alpha$-$(ET)_2I_3$, interlayer SOC leads to a 3D Dirac semimetal state when SIS is preserved [13], while SIS breaking in the interlayer transfers results in a spin-degenerate Weyl semimetal state. In $\alpha$-$(BETS)_2I_3$, we demonstrated that a 3D weak TI state is realized under the presence of interlayer SOC and SIS, as confirmed using the parity criterion [14].

In this paper, we investigate these 3D topological states by analyzing the Wannier charge centers (WCCs) [15-18] in 3D $\alpha$-type conductors. While this method has often been used to evaluate the topological $Z_2$ numbers in TIs, we apply it also to topological semimetals. It is a rather intuitive approach and has the advantage of being applicable even to systems without SIS. The WCC $<y>$ is defined as the expected value of the position of the hybrid Wannier function, which is obtained by Fourier transforming the Bloch functions in only one direction (assumed to be the $y$-direction here). It can be calculated from the Berry phase along the $k_y$-direction across the Brillouin zone (BZ) with fixed $k_x$ and $k_z$. In the case of degenerated bands, the Berry phases are obtained as Wilson loop eigenvalues in the non-Abelian formalism. When the other wave number ($k_x$, $k_z$) change adiabatically, the WCCs may shift, implying a shift of wave functions that can contribute to the topological transport of charge and/or spin. We study the WCC $<y>$ as a function of ($k_x$, $k_z$ ) for each possible 3D topological phase in $\alpha$-type organic conductors.

The 3D tight-binding model for $\alpha$-type organic conductors, $\alpha$-$(ET)_2I_3$ and $\alpha$-$(BETS)_2I_3$, was employed following Ref. [12]. In this model, the interlayer hoppings $t_A$, $t_{A'}$, $t_B$, and $t_C$ between equivalent sites on adjacent layers are added to the conventional 2D tight-binding model [1], which preserves SIS. In-plane SOC $\lambda$ and interlayer SOC $\lambda$' are additionally introduced to reflect the configuration of the $I_3^-$ anions [12, 19]. In the 2D model without interlayer coupling, the system becomes a massless DF system corresponding to $\alpha$-$(ET)_2I_3$ when the in-plane SOC is negligible, whereas it becomes a TI corresponding to $\alpha$-$(BETS)_2I_3$ when the in-plane SOC is finite.

We first examine the WCCs of multilayered $\alpha$-type organic conductors in the absence of in-plane SOC ($\lambda$=0), corresponding to a model for 3D $\alpha$-$(ET)_2I_3$. Following our previous work [12], we study four cases with or without interlayer SOC and with or without SIS breaking in the interlayer hopping. Figure 1 shows an example of the band structure in the Dirac semimetal case with finite interlayer SOC and no SIS breaking. In the in-plane dispersion at $k_z = 0$ (Fig. 1(a)), the spin-degenerate valence band $E_3(\mathbf{k})$ and conduction band $E_4(\mathbf{k})$ touch at two Dirac points, which extend into nodal lines along the interlayer direction. The interlayer SOC opens gaps along these nodal lines, leaving band-contact points at $k_z =$ 0 and $\pm\pi/c$, where $c$ is the interlayer lattice constant, as shown in Fig. 1(b) [12].

Figure 2 illustrates the interlayer band dispersion and the WCC sheets, i.e., $\langle y \rangle$ as functions of $(k_x, k_z)$, for each case. In these WCC panels, the gray planes at $\langle y \rangle/a = 0, \pm1$ ($a$: lattice constant) indicate the centers of the unit cells (lattice points) along the $y$-direction. When 2D DF layers are simply stacked while preserving SIS and without SOC, the multilayer system becomes a 3D nodal-line semimetal with two straight nodal lines parallel to the $k_z$-axis. The corresponding WCC sheets are flat but discontinuously cut at the singular nodal lines, as shown in Fig. 2(a). In this case, electrons moving on a flat WCC sheet acquire no anomalous velocity in the $y$-direction. When the interlayer SOC becomes finite ($\lambda' \neq 0$), band gaps open along the nodal line, leaving band-contact points at $k_z = 0$ and $\pm\pi/c$, as already seen in Fig. 1. Although the band structure retains twofold spin degeneracy, the WCC sheets exhibit spin splitting and are continuous except at the contact points, as observed in Fig. 2(b). Each pair of split WCC sheets winds in opposite directions around the contact points and connects adjacent unit cells in the $y$-direction, forming double spirals, as discussed below. An enlarged view of the spiral structure of the WCC sheets within the white box is shown in Fig. 3(d). This system is a type-I Dirac semimetal, featuring Dirac points where spin-degenerate subbands cross with opposite chirality [12]. Such a state can be realized under both SIS and TRS [13].

On the other hand, when SIS is broken by introducing nonuniform interlayer hoppings ($t_A \neq t_{A'}$) without SOC, the spin-degenerate conduction and valence bands have two band-contact points at $k_z = \pm\pi/2c$ along the nodal lines, as shown in Fig. 2(c). This system corresponds to a spinless type-II Weyl semimetal [12], expected in the absence of SIS [13]. The corresponding WCC sheets are also spin-degenerate, and wind around a Weyl point in the same directions, forming a single twofold spiral, as observed in the Fig. 2(c). These spirals continuously connect the adjacent unit cell in the $y$-direction.

Moreover, when the interlayer SOC becomes finite in addition to the SIS breaking, the energy bands exhibit zero-field spin splitting, and each Weyl point splits into two nondegenerate Weyl points with the same chirality, as seen in Fig. 2(d). This configuration corresponds to a type-II Weyl semimetal [12]. The WCC sheets are spin-split, and each sheet winds around a Weyl point, forming a single spiral, as also shown in Fig. 2(d). An enlarged view of the WCC spiral structure within the white box is shown in Fig. 3(c).

Although the above features of organic Dirac/Weyl semimetals were previously discussed in terms of Berry curvature [12], chiral transport phenomena in Dirac/Weyl semimetals can be understood more intuitively using the WCC picture. The Dirac/Weyl points, where the conduction and valence bands linearly touch, are singular points at which the WCC is undefined. At a Weyl point with a given chirality, the WCC is discontinuous and typically forms a spiral structure connecting neighboring unit cells along the $y$-direction. This occurs because the WCC configuration must return to its original value after completing a loop around the Weyl point. This situation is schematically illustrated in Fig. 3(a). The winding direction of the spiral WCC corresponds to the chirality of the Weyl point. In contrast, for Dirac semimetals, two spin-degenerate subbands produce oppositely winding WCC sheets that interlace around the Dirac point, as shown in Fig. 3(b). These spiral WCC structure around Dirac/Weyl points are clearly seen in Fig. 2(b), (c), and (d).

When the electron undergoes adiabatic circular motion in the ($k_x$, $k_z$)-plane, such as the projection of the cyclotron orbital motion on the Fermi surface under a magnetic field applied in the $y$-direction, the WCC moves upwards or downward along the spiral sheets to the neighboring unit cell. This process is induced by the adiabatic deformation of Wannier functions and pumps current along the $y$-direction without an electric field or dissipation. This offers a WCC-based interpretation of the CME. In fact, one complete cycle in the ($k_x$, $k_z$)-plane causes WCC $\langle y \rangle$ to shift by an integer $N$ multiple of the lattice constant $a$ in the $y$-direction. This gives rise to the standard expression for the CME: $j_y^{(\mathrm{CME})} = -(e^2/h^2)N(\mu - E_0)B_y$, where $\mu$ and $E_0$ are the chemical potential and a reference energy, respectively [10]. The integer $N$ corresponds to the monopole charge, or the Chern number of the closed surface enclosing the Weyl or Dirac point, and reflects the chirality, which is sgn($N$) in the valence band and −sgn($N$) in the conduction band. Under an electric field $E_y$ in the $y$-direction, the CME current does work $j_y^{(\mathrm{CME})}E_y$, leading to a shift in the chemical potential $\mu$ around a single Weyl point, depending on sgn($N$). The resulting imbalance in chemical potential between Weyl points of opposite chirality give rise to a net CME current along the magnetic field direction, which manifests as negative longitudinal magnetoresistance or the planer Hall effect. This topological transport originates from the adiabatic shift of Wannier functions and provides an additional contribution to the conventional magnetotransport.

The spiral structures around the Dirac/Weyl points in Fig. 2(b), (c), and (d) can give rise to the CME, provided that the Dirac/Weyl points lie within closed cyclotron orbits in the ($k_x$, $k_z$)-plane. As seen in the interlayer band dispersions in Fig. 2, electron and hole pockets appear centering at $k_z = 0$ and $\pm\pi/c$, tracing the original nodal line. For the CME to occur, the Dirac/Weyl points must lie within the Fermi surface of these pockets. Among the topological semimetal states of $\alpha$-$(ET)_2I_3$ shown in Fig. 1, the type-I Dirac semimetal state appears the most plausible to explain the experimentally observed CME-related transport phenomena [8] under in-plane magnetic fields. In the type-II Weyl semimetal states, Weyl points locate outside or near edge of Fermi surfaces, so that it is difficult to induce CME under in-plane magnetic fields.

Prior to experiments on CME-related transport phenomena, a type-II spin-degenerate Weyl semimetal state with both TRS breaking and SIS breaking was theoretically predicted as a flux state induced by electron exchange interactions with multiple interlayer transfers [20, 21], although it was referred to as a "Dirac semimetal" in Ref. [20]. This flux state lies outside the scope of the present discussion under TRS. However, the CME-related transport (such as the planar Hall effect) in this state, which was also discussed in Ref. [22], is expected to be weak because of its type-II dispersion with Weyl points at $k_z = \pm\pi/2c$ [20]. In fact, for type-II Weyl dispersions tilted in the $k_z$-direction, it has been generally argued that the CME vanishes over a finite range of magnetic-field orientations, including the in-plane directions (perpendicular to $k_z$) [23].

Next, we investigate the WCCs of $\alpha$-type organic conductors with finite in-plane SOC ($\lambda \neq 0$), corresponding to a model of multilayered $\alpha$-$(BETS)_2I_3$. When the interlayer hopping is negligible, the system becomes a 2D TI due to the in-plane SOC [19]. Here, we examine the effects of interlayer hopping under the same conditions as in the $\lambda = 0$ case: with or without interlayer SOC, and with or without SIS breaking.

Figure 4 shows the interlayer dispersion and the corresponding WCC sheets for the case in which finite interlayer SOC ($\lambda' \neq 0$) and SIS breaking (e.g., $t_A \neq t_{A'}$) are introduced in the interlayer coupling. The WCC sheets exhibit spin splitting but remain degenerate at the time-reversal invariant momenta (TRIMs), forming Kramers pairs. Each sheet switches its pairing partner at the TRIMs $k_x = 0, \pm\pi/b$ along the $k_x$-direction, but no such switching occurs along the $k_z$-direction. Furthermore, as $k_x$ adiabatically changes from $-\pi/b$ to $\pi/b$ across the BZ at fixed $k_z$, the split WCCs $\langle y \rangle$ shift by one lattice constant $a$ in opposite directions. In contrast, no such shift occurs in the $k_z$-direction. These features are indicative of a 3D weak TI. In the other cases, whether with or without interlayer SOC and with or without SIS breaking, the above characteristics remain unchanged. Therefore, all of these cases correspond to 3D weak TI states.

The switching of WCC partners at the neighboring TRIMs in a time-reversal invariant plane in the 3D BZ implies a nontrivial $Z_2$ index $\nu = 1$ for that plane. From Fig. 4(b), we can conclude $\nu_0 = 1+1 = 0$ (mod 2), $\nu_x = 0$, and $\nu_z = 1$, since $\nu = 1$ on the planes $k_z = 0$ and $k_z = \pi/c$. Finally, the topological $Z_2$ indices are determined from the WCC flow as $(\nu_0; \nu_x, \nu_y, \nu_z) = (0; 0, 0, 1)$ for all cases. In our previous work [12], we identified the weak TI state only in the presence of SIS, using Fu's parity criterion [14], which cannot be applied when SIS is broken. In contrast, the WCC method is broadly applicable regardless of whether SIS is preserved or broken.

These results are consistent with the fact that no significant topological change in the band structure occurs in the present conditions. The 3D strong TI state, which was proposed based on the experimental observation of surface transport [11], is unlikely to be realized in $\alpha$-(BETS)$_2$I$_3$ without a topological phase transition involving a band structure modification.

In conclusion, we have demonstrated the characteristic features of WCCs for the possible 3D topological phases expected in $\alpha$-type organic conductors $\alpha$-(ET)$_2$I$_3$ and $\alpha$-(BETS)$_2$I$_3$ at low temperatures, where interlayer hopping cannot be neglected. For the 3D topological semimetal states expected in multilayered $\alpha$-(ET)$_2$I$_3$ with vanishing in-plane SOC, the WCC sheets exhibit spiral structures around the Dirac/Weyl points, leading to the CME under in-plane magnetic fields. Considering the relative position of nodal points and Femi surfaces, the type-I Dirac semimetal state induced by interlayer SOC appears to be the most plausible explanation for the experimentally observed transport phenomena. On the other hand, in the 3D TI states expected in multilayered $\alpha$-(BETS)$_2$I$_3$ with finite in-plane SOC, the WCC sheets consistently shows the characteristics of a weak TI, with a topological $Z_2$ index set of (0; 0, 0, 1), regardless of the presence of interlayer SOC and/or SIS breaking in the interlayer hopping. The strong TI state suggested by experiments appears unlikely to be realized without a topological transition accompanied by a band structure modification.

**Acknowledgements**

This work was partially supported by JSPS KAKENHI Grant Numbers JP23K03297 and JP24H01610. The author thanks Prof. Tatsuo Hasegawa and Prof. Yasuhiro H. Matsuda for providing an excellent research environment.

**References**

*osada@issp.u-tokyo.ac.jp

**Figure 1** (Osada)

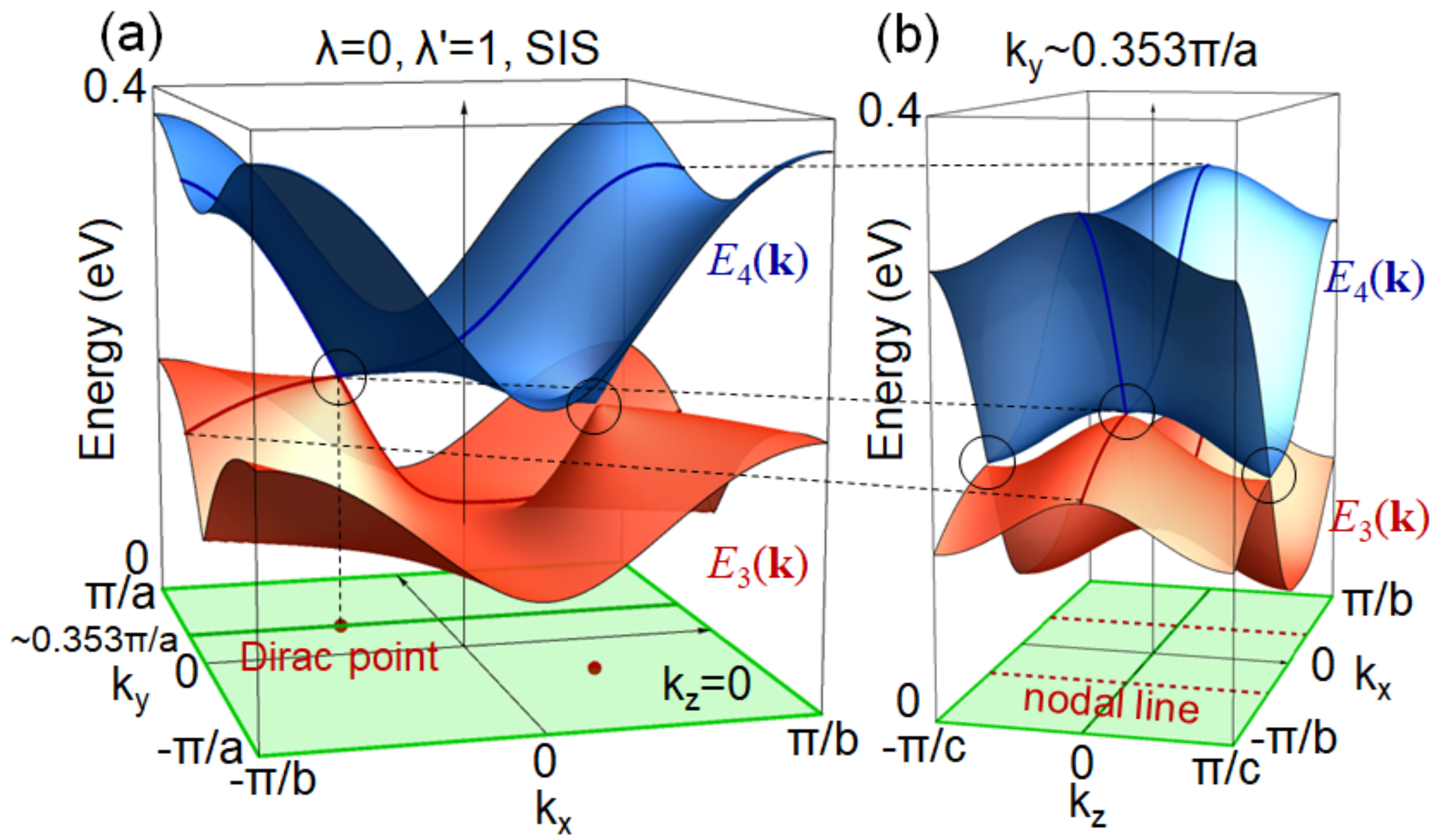

**FIG. 1.** (color online)

Dispersion of the conduction and valence bands in the Dirac semimetal phase of multilayered $\alpha$-$(ET)_2I_3$ with interlayer SOC. Both bands exhibit twofold spin degeneracy. All interlayer transfer energies were set to 10meV. (a) In-plane ($k_x$, $k_y$) dispersion at $k_z = 0$. (b) Interlayer ($k_z$, $k_x$) dispersion along a nodal line at $k_y \sim 0.353\pi/a$. The Dirac points are indicated by circles.

**Figure 2** (Osada)

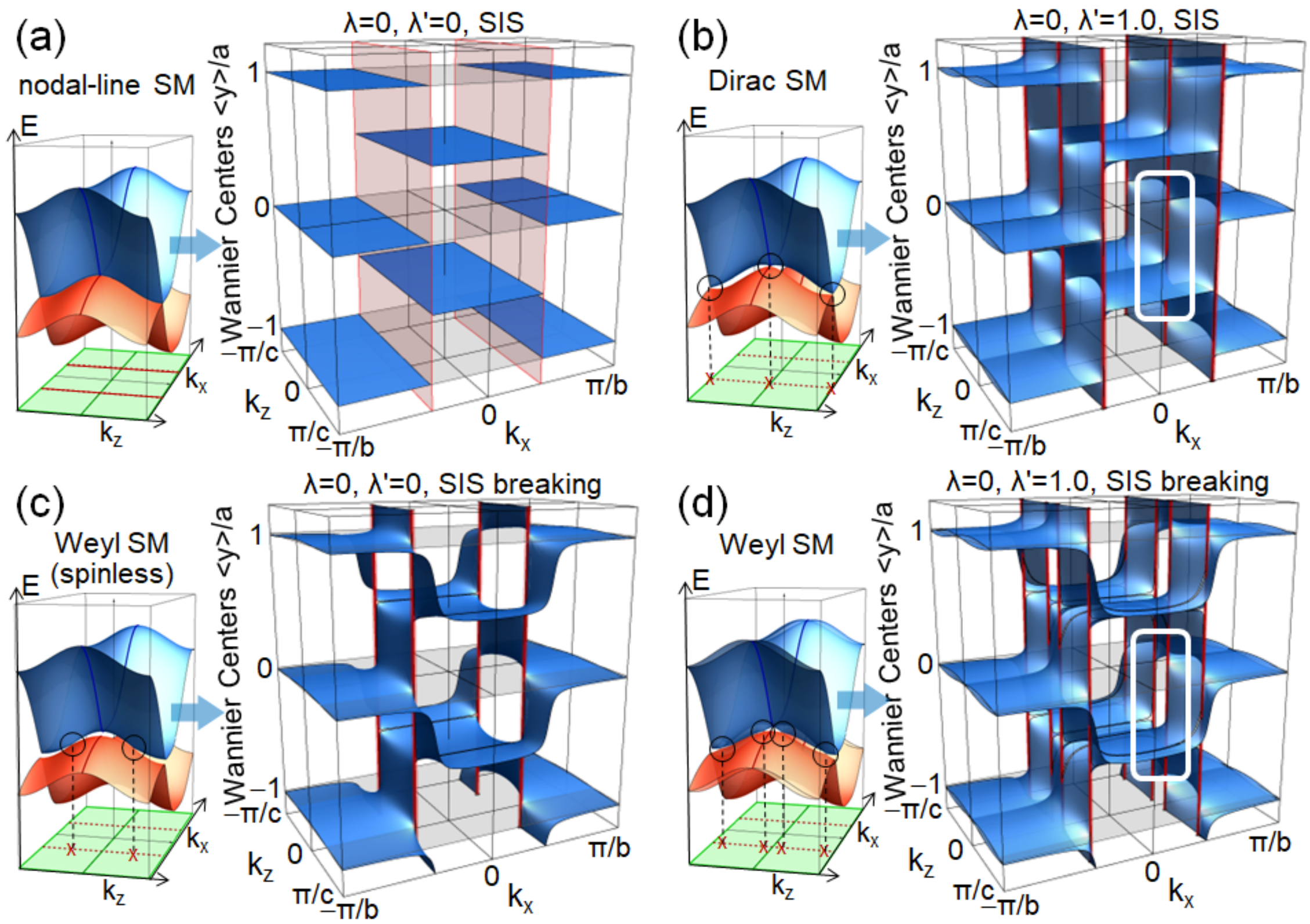

**FIG. 2.** (color online)

Interlayer dispersion and corresponding WCC sheets of the conduction band in four possible 3D topological semimetal states of multilayered $\alpha$-$(ET)_2I_3$. (a) Nodal-line semimetal of simple stacking with SIS preserved. (b) Dirac semimetal with SIS preserved, induced by interlayer SOC. (c) Spin-degenerate Weyl semimetal induced by SIS breaking in interlayer coupling. (d) Weyl semimetal induced by both interlayer SOC and SIS breaking. In the WCC panels, red planes and red lines indicate the singular lines (nodal lines) or points (Dirac/Weyl points), respectively. $\lambda$ and $\lambda$' denote the in-plane and interlayer SOC parameters, respectively [12]. Enlarged views of the boxed regions in (b) and (c) are shown in Fig. 3(d) and Fig. 3(c), respectively.

**Figure 3** (Osada)

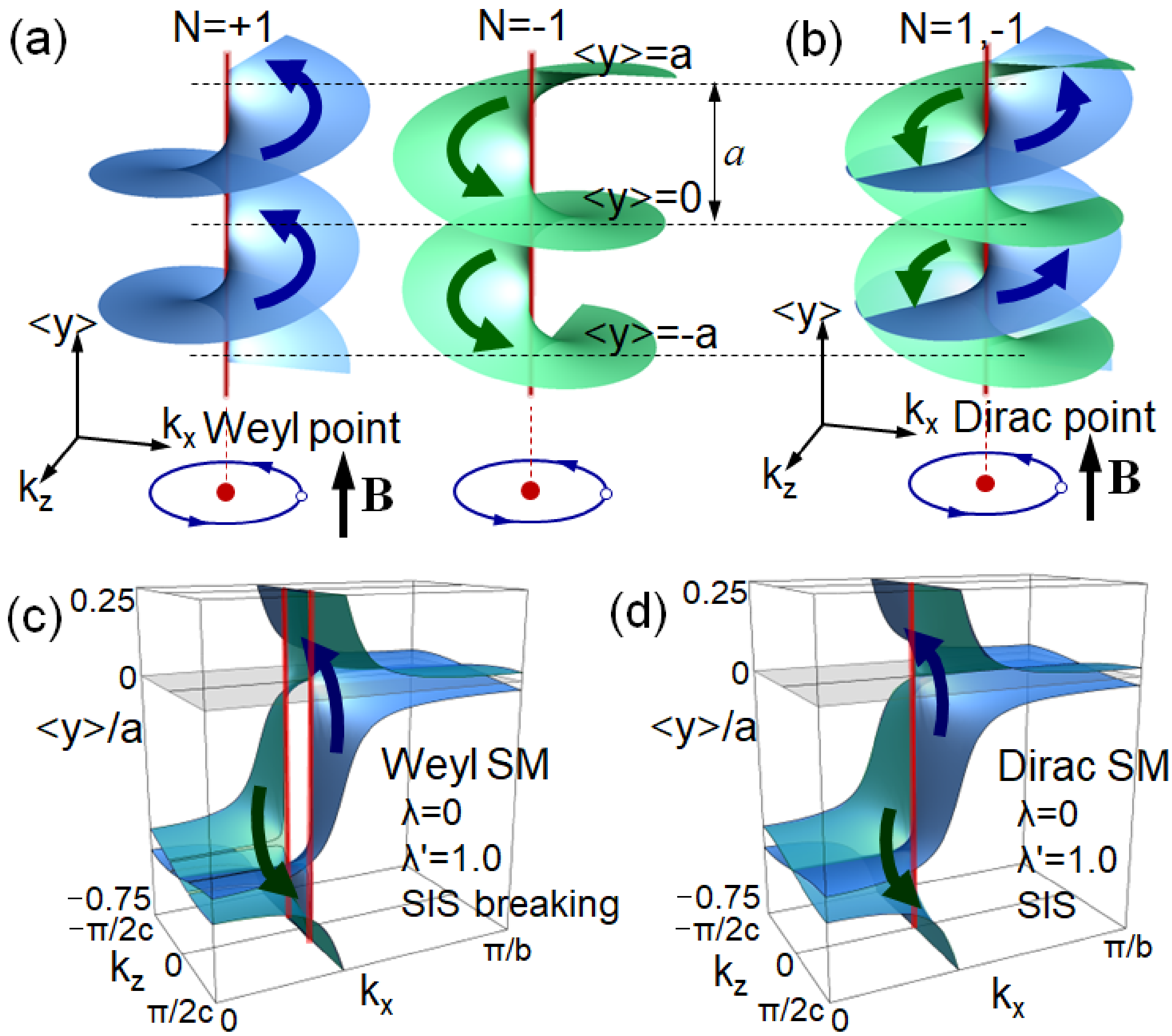

**FIG. 3.** (color online)

Schematics of spiral WCC sheets and the CME in Dirac/Weyl semimetals. Red lines indicate Dirac/Weyl points. (a) Spiral WCC sheets around Weyl points with opposite monopole charges $N$ (with chirality of $-\mathrm{sgn}(N)$ for the conduction band). When an electron completes one cyclotron motion around a Weyl point in the $(k_x,k_y)$-plane, it is transported by one lattice constant $a$ in the $y$-direction. (b) Two WCC sheets with opposite chirality wind around a Dirac point. (c)(d) Details of WCC spiral structures for the Weyl and Dirac semimetal states in $\alpha$-$(ET)_2I_3$, corresponding to Fig. 2(d) and 2(b), respectively.

**Figure 4** (Osada)

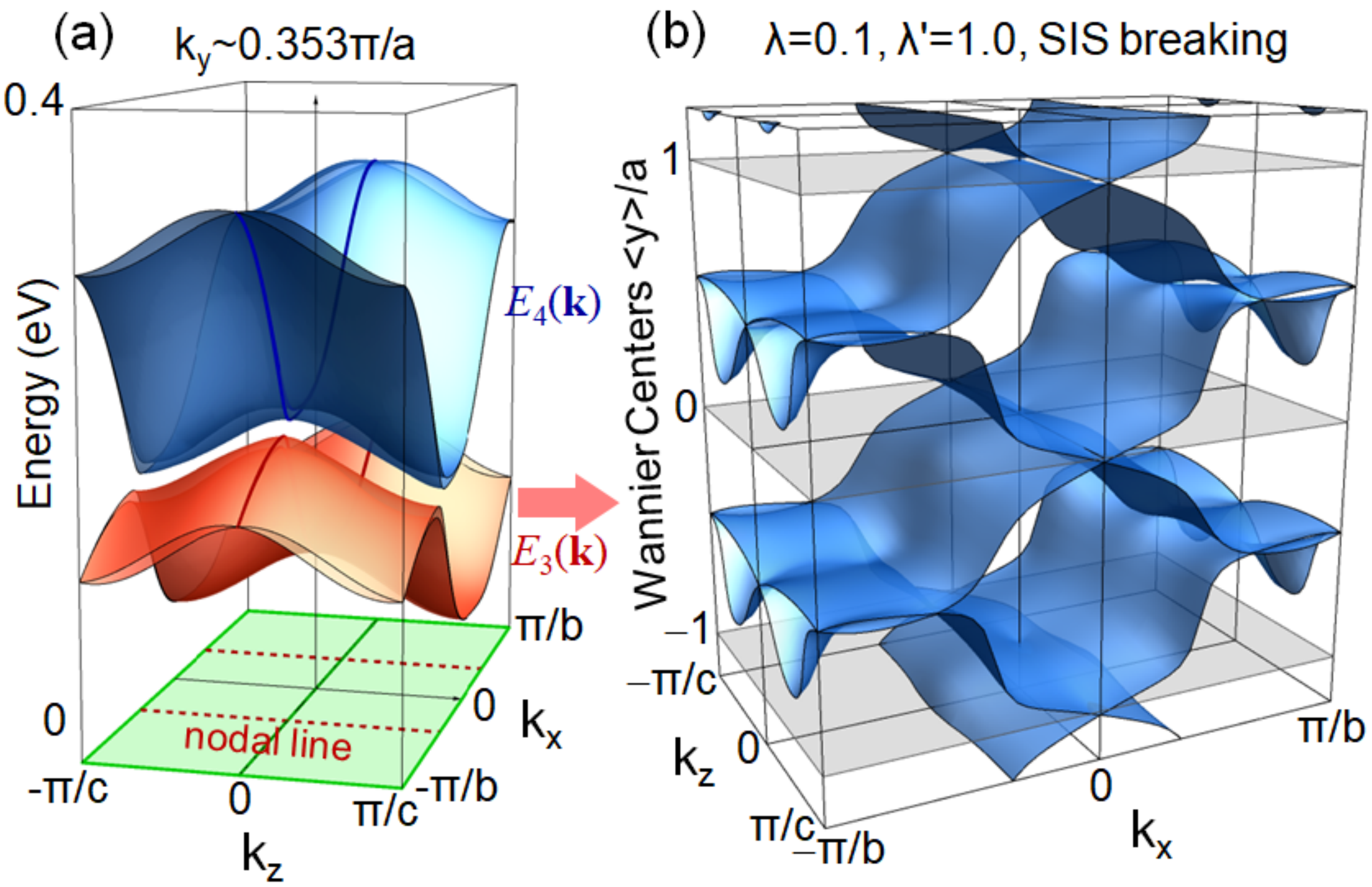

**FIG. 4.** (color online)

(a) Interlayer band dispersion in a topological insulator state of multilayered $\alpha$-(BETS)$_2$I$_3$ with in-plane and interlayer SOC, and SIS breaking in the interlayer coupling. Spin splitting is observed at zero magnetic field. (b) Corresponding WCC sheets of the valence band. The WCCs also exhibit spin splitting under SOC. The $Z_2$ indices $\nu_0 = 0$, $\nu_x = 0$, and $\nu_z = 1$ are obtained.